\documentclass[twocolumn,showpacs,preprintnumbers,amsmath,amssymb,prl,superscriptaddress]{revtex4}

\usepackage{amsfonts}
\usepackage{graphicx}
\usepackage{dcolumn}
\usepackage{bm}
\usepackage{color}

\begin{document}

\preprint{APS/123-QED}

\title{Verification of the Wiedemann-Franz law in YbRh$_2$Si$_2$ at a quantum critical point}

\author{Y. Machida}
\author{K. Tomokuni}
\author{K. Izawa}
\affiliation{Department of Physics, Tokyo Institute of Technology, Meguro 152-8551, Japan}

\author{G. Lapertot}
\author{G. Knebel}
\author{J.-P. Brison}
\author{J. Flouquet}
\affiliation{SPSMS, UMR-E CEA / UJF-Grenoble 1, INAC, Grenoble, F-38054, France}

\date{\today}

\begin{abstract}
The thermal conductivity measurements have been performed on the heavy-fermion compound YbRh$_2$Si$_2$
down to 0.04 K and under magnetic fields through a quantum critical point (QCP) at $B_{\rm c}$ = 0.66 T $\parallel$ $c$-axis. 
In the limit as $T\rightarrow$ 0, we find that the Wiedemann-Franz law is
satisfied within experimental error at the QCP despite the destruction of the standard signature of Fermi liquid.
Our results place strong constraints on models that attempt to describe the nature of  unconventional quantum criticality of YbRh$_2$Si$_2$.
\end{abstract}


\maketitle
A quantum critical point (QCP) defines a second-order transition at zero temperature that is driven by external parameters, 
such as pressure, magnetic field, and chemical substitution~\cite{stewart,lohneysen_review,gegenwart_review}. 
Near the QCP, the Fermi liquid (FL) behavior is destroyed by diverging quantum fluctuations,
and in consequence anomalous properties quoted as non-Fermi liquid (NFL) behaviors show up. 
Given that a  number of intriguing phenomena, 
e.g., unconventional superconductivity~\cite{mathur}, exotic electronic states~\cite{borzi}, and nontrivial spin states~\cite{MnSi}, are found in the vicinity of the QCP, 
it is essential to understand the nature of the quantum criticality. 

A key element in the debate about the quantum criticality in heavy fermion metals is concerning whether the quantum critical physics can
be understood by the conventional spin-fluctuation theories~\cite{millis,hertz,moriya},
or whether a new framework, invoking the critical breakdown of the Kondo effect at the QCP is required~\cite{si,coleman}.
A major difference of  the two scenarios relies on quite different fates for the FL state. 
In the conventional scenario~\cite{millis,hertz,moriya}, 
fluctuations are concentrated at hot spots on the Fermi surface (FS), so that
the FL state is retained on part of the FS at the QCP. 
In the Kondo breakdown scenario~\cite{si,coleman}, by contrast,
fluctuations are thought to cover the entire FS, 
leading to a strong breakdown of the quasiparticle picture~\cite{si,coleman}. 
What is desperately needed for differentiating between the two scenarios is therefore to conclusively determine whether the heavy quasiparticles survive at the QCP.

A crucial test of this issue is a verification of the Wiedemann-Franz law at the QCP, 
which states that the ratio of the thermal conductivity $\kappa$ to the electrical conductivity $\sigma$
is a universal constant in the $T\rightarrow$ 0 limit: $\kappa/\sigma T\equiv L = L_0$,
where $L_0=\frac{\pi^2}{3}(\frac{k_{\rm B}}{e})^2=2.44\times 10^{-8}$ W$\Omega$/K$^2$ is the Sommerfeld value. 
A violation of this law would imply a profound breakdown of the FL theory~\cite{senthil,colemanWF,podolsky,pepin}.
Experimentally, however, the Wiedemann-Franz law appears to be universal as $T\rightarrow0$ and
no material has been reported to violate this law up to date. 
(We note that a deviation from the Wiedemann-Franz law has been reported in CeCoIn$_5$~\cite{tanatar} as we will discuss later).

In the Letter, 
we have chosen to study the Wiedemann-Franz law in YbRh$_2$Si$_2$, a tetragonal heavy fermion compound~\cite{trovarelli}.
YbRh$_2$Si$_2$ has provided a rare opportunity to probe the electronic properties
near the QCP by using the magnetic field $B$ as a tuning parameter. 
The very weak antiferromagnetic (AF) order ($T_{\rm N}\sim$ 0.07 K) is suppressed by a small magnetic field ($B_{\rm c}$ = 0.66 T $\parallel$ $c$-axis),
giving rise to the NFL behaviors 
characterized by a divergence in transport and thermodynamic quantities~\cite{gegenwart}. 
These anomalous critical phenomena, which do not follow the conventional theories, have been interpreted as observing the breakup of the 
quasiparticles at the QCP~\cite{coleman,custers}.
According to this interpretation, one would speculate a violation of the Wiedemann-Franz law in YbRh$_2$Si$_2$.
Indeed, several theories of the unconventional quantum criticality predict that it may be violated at the QCP~\cite{senthil,colemanWF,podolsky,pepin}.
Furthermore, a new temperature scale of $T^*$ has been identified in the $T$-$B$ phase diagram.
Across this scale, the Hall coefficient exhibits a substantial change as a function of the magnetic field 
as a possible signature of a breakdown of the Kondo effect at the QCP involving an abrupt change of the FS volume~\cite{paschen}.
To test such a scenario, we have performed a set of thermal and charge conductivity measurements on YbRh$_2$Si$_2$,
which demonstrate a verification of the Wiedemann-Franz law in the limit as $T\rightarrow$ 0
at the QCP, implying that there is no breakdown of the quasiparticles.

A high quality single crystal of YbRh$_2$Si$_2$ with low residual resistivity $\rho_0 \approx$ 0.9 $\mu\Omega$cm and a high residual resistivity ratio $\sim$ 90
was prepared using the indium flux~\cite{knebel}.
The thermal conductivity $\kappa$ was measured
by employing the one-heater-two-thermometers steady-state method in a dilution refrigerator. 
The heat current $q$ was injected parallel to
the $ab$-plane, and the magnetic field $B$
was applied parallel to the $c$-axis on the sample with a size of
$1.6 \times 2.2 \times 0.05$ mm$^3$.
The thermal contacts with resistance of $\leq$10 m$\Omega$ at room temperature
were made by using a spot welding technique.
The same contacts were used to measure the resistivity $\rho$ by a standard four-contact method.
We note that $\rho_0$ of our sample is lower than the one ($\rho_0 \approx$ 1.5 $\mu\Omega$cm)
of the previous report with $B$ $\parallel$ $ab$~\cite{pfau},
which provides good condition for enhancing the electronic contributions with respect to an additional contribution,
such as magnon.

\begin{figure}[t]
\begin{center}
\includegraphics[scale =0.7]{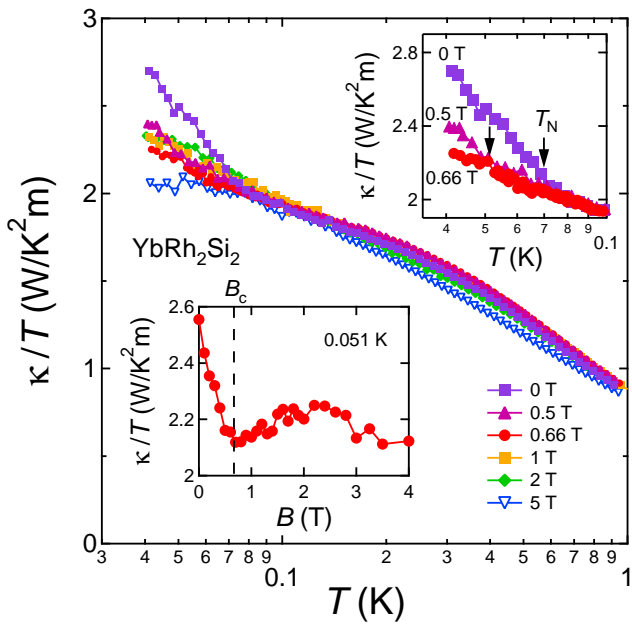}
\end{center}
\vspace{-0.5cm}
\caption{\label{fig.1} (color online). Temperature dependence of the thermal conductivity divided by temperature $\kappa(T)/T$ of YbRh$_2$Si$_2$
at various magnetic fields with the enlarged plot around the
N$\acute{\rm e}$el ordering temperature $T_{\rm N}$ (upper inset), in semilogarithmic plots. 
Lower inset: Field dependence of $\kappa(B)/T$ at 0.051 K. $B_{\rm c}$ = 0.66 T denotes the critical field.}
\end{figure}

Figure~\ref{fig.1} displays the temperature dependence of the thermal conductivity divided by temperature $\kappa(T)/T$
at various magnetic fields.
Here, we note that the phonon contribution to the measured $\kappa$ is negligible within the temperature range of interest,
which is estimated to be at most $\approx$ 0.2 $\%$ of $\kappa$ at 0.1 K via the phonon specific heat~\cite{hartmann2} with an assumption of boundary scattering limited conduction.
Under zero field, we find a remarkable upturn in $\kappa(T)/T$
below the N$\acute{\rm e}$el ordering temperature $T_{\rm N}\approx$ 0.07 K.
At 0.5 T, $T_{\rm N}$ is reduced to $\sim$ 0.05 K,
and it entirely disappears at the critical field $B_{\rm c}$ = 0.66 T (the upper inset of Fig.~\ref{fig.1}).
Even above $B_{\rm c}$, $\kappa(T)/T$ shows a steep increase down to the lowest temperature, but it is less sensitive to the fields up to 3 T.
A FL-like saturation behavior is eventually observed below about 0.06 K at 5 T. 
The lower inset shows the field dependence of $\kappa(B)/T$ measured at 0.051 K.
We find a rapid decrease of $\kappa(B)/T$ up to $B_{\rm c}$, indicating that the N$\acute{\rm e}$el ordering is indeed suppressed right at $B_{\rm c}$.

\begin{figure}[t]
\begin{center}
\includegraphics[scale =0.6]{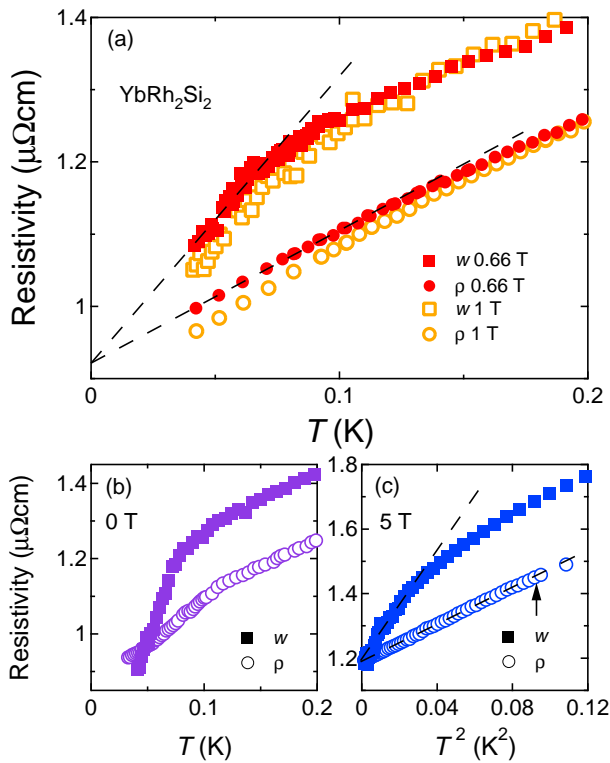}
\end{center}
\vspace{-0.5cm}
\caption{\label{fig.2} (color online). Temperature dependence of the heat resistivity $w(T)$ and the charge resistivity $\rho(T)$ of YbRh$_2$Si$_2$ at
(a) $B_{\rm c}$ = 0.66 T and 1 T and at (b) 0 T, plotted vs $T$, and at (c) 5 T, plotted vs $T^2$, respectively.
The dashed lines represent linear fits to the data.
An arrow in (c) indicates the upper bound of $T^2$ dependence of $\rho(T)$.}
\end{figure}

Figure~\ref{fig.2} shows the temperature dependence of the heat resistivity $w(T)=L_0(T/\kappa)$ and the charge resistivity $\rho(T)$ 
at (a) $B_{\rm c}$ = 0.66 T and 1 T, (b) 0 T, and (c) 5 T, respectively.
First of all, one may immediately notice that $w(T)$ is larger than $\rho(T)$  in a wide temperature and field range,
indicating that $w$ is more strongly influenced by the scattering than $\rho$.
This is due to an excess scattering process affecting only the heat conduction, in which the energy of the conduction electron is changed.
A source of inelastic scattering is possibly magnetic fluctuations enhanced by the proximity to the QCP.
On cooling at $B$ = 0 T, $\rho(T)$ linearly decreases with temperature down to $\sim$ 0.1 K and substantially drops
at $T_{\rm N}$.
A sharp decrease is also observed in $w(T)$ at $T_{\rm N}$ followed by a $T$-linear dependence above $\sim$ 0.1 K.
At temperatures well below $T_{\rm N}$, $w(T)$ becomes less than $\rho(T)$, indicating the presence of extra heat carrier which may
attribute to antiferomagnetic magnons as concluded in the previous report~\cite{pfau}.

In the quantum critical regime, at $B = B_{\rm c}$, $\rho(T)$ varies $T$-linearly below 0.12 K down to the lowest temperature.
$w(T)$ also exhibits the linear $T$-dependence, but in the limited temperature range ($\sim$ 0.1 K $< T <$ $\sim$ 0.2 K), and 
shows a downturn below $\sim$ 0.1 K as it becomes closer to $\rho(T)$
to satisfy the Wiedemann-Franz law.
The downturn was also observed in the previous report, but it was attributed to
overdamped magnons which exist for antiferromagnetically ordered materials substantially above $T_{\rm N}$~\cite{pfau}.
With this interpretation, the low-temperature part of $w(T)$ below the downturn was excluded from the analysis,
and a residual value $w_0$ was determined by the linear extrapolation of high-temperature $w(T)$.
This gives rise to a discrepancy between $w_0$ and $\rho_0$ ($\rho_0 < w_0$),
leading a violation of the Wiedemann-Franz law ($L/L_0 \approx 0.9$)~\cite{pfau}.
We should emphasize that this conclusion is problematic in the analysis of $w(T)$ because
1) the discrepancy ($\rho_0 < w_0$) is a natural consequence as $w_0$ is obtained by a fitting to the high-temperature data
where heat carriers are subject to strong inelastic scattering,
2) the downturn does not originate from the magnons because a similar feature is found even
away from the antiferromagnetic phase; $w(T)$ also exhibits the downturn at 1 T as shown in Fig.~\ref{fig.2}(a),
where the system becomes FL below $\sim$ 0.05 K~\cite{gegenwart}.
3) if the magnon contribution is the case, $w(T)$ should rise to hold $L/L_0 \approx 0.9$ at low temperatures 
since the magnetic contribution will vanish in the $T$ = 0 limit.
Clearly, there is no such behavior in our data as well as in the previous one~\cite{pfau}.
We stress that the behavior of $w(T)$ found at $B_{\rm c}$ is the same as the previous one~\cite{pfau} irrespective of the difference in the applied field directions, and 
the main questionable point is the analysis.
Rather, the downturn can be interpreted as the contribution of purely electronic excitations as we will discuss in more detail.
On the basis of these considerations, the downturn is seriously taken into account for our analysis.
By assuming the $T$-linear dependence of $w(T)$ below about 0.07 K,
both $w(T)$ and $\rho(T)$ are extrapolated to $T$ = 0 as denoted by the dashed lines.
Remarkably, the two lines converge at $T$ = 0,
i.e., $L/L_0 = \rho_0/w_0$ = 1, leading a verification of the Wiedemann-Franz law at the QCP.

In the FL state at $B$ = 5 T, 
both $w(T)$ and $\rho(T)$ distinctly display the $T^2$ dependence (Fig.~\ref{fig.2}(c)).
The extrapolations denoted by the dashed lines show that $w_0 = \rho_0$ at $T$ = 0.
The upper limit of $\rho(T) \propto T^2$ defined as an energy scale of the FL, $T_{\rm FL}$ is denoted by an arrow.
Using the extrapolated values $w_0$ and $\rho_0$, the field variation of the Lorenz ratio $L/L_0=\rho_0/w_0$ in the zero temperature limit 
is given in Fig.~\ref{fig.4} for $B \geq B_{\rm c}$.
It can be clearly seen that $L/L_0$ is unity within the experimental error for $B \geq B_{\rm c}$, indicating that
the Wiedemann-Franz law is obeyed not only in the FL state but also at the QCP.

\begin{figure}[t] 
\begin{center}
\includegraphics[scale =0.65]{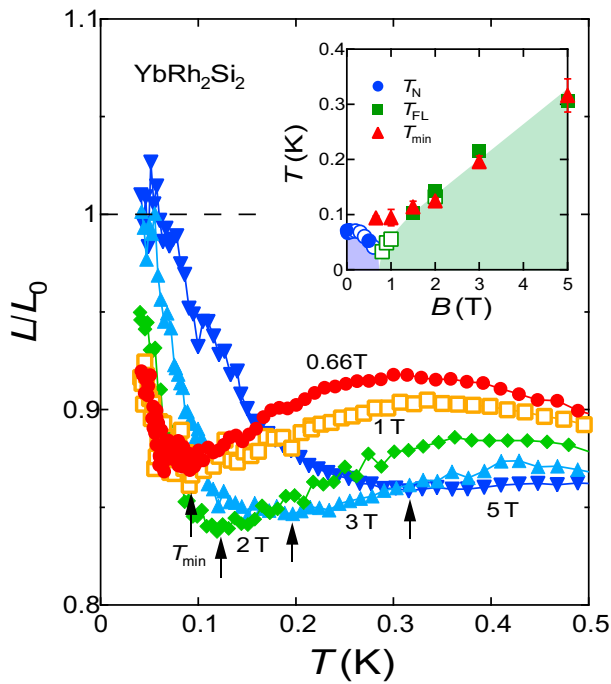}
\end{center}
\vspace{-0.5cm}
\caption{\label{fig.3} (color online). Temperature dependence of the Lorenz ratio $L(T)/L_0$ of YbRh$_2$Si$_2$ above the critical field $B_{\rm c}$ = 0.66 T.
Arrows denote minima $T_{\rm min}$ in $L(T)/L_0$ at each field.
Inset: $B$-$T$ phase diagram of YbRh$_2$Si$_2$, including 
the N$\acute{\rm e}$el ordering temperature $T_{\rm N}$, 
the upper bound of $T^2$ electrical resistivity $T_{\rm FL}$, and
the positions of minimum $T_{\rm min}$ of the Lorenz ratio $L(T)/L_0$, respectively,
together with $T_{\rm N}$ and $T_{\rm FL}$ (open symbols) from Ref.~\cite{gegenwart}.}
\end{figure}

\begin{figure}[t]
\begin{center}
\includegraphics[scale =0.6]{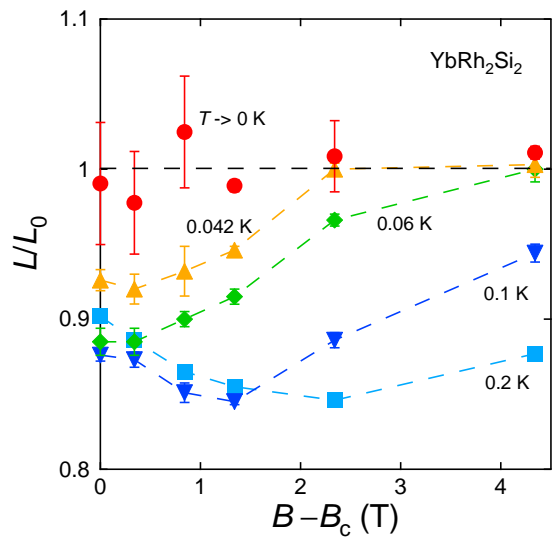}
\end{center}
\vspace{-0.5cm}
\caption{\label{fig.4} (color online). Field dependence of the Lorenz ratio $L(B)/L_0$ of YbRh$_2$Si$_2$ in the zero temperature limit
and at finite temperatures.}
\end{figure}

Next, to shed light on the effect of the critical fluctuations influence on the quasiparticle scattering, 
we focus on the Lorenz ratio $L(T)/L_0 = \rho(T)(\kappa(T)/T)/L_0$ at finite temperature.
Figure~\ref{fig.3} shows the temperature variation of $L(T)/L_0$ under several fields.
At $B = B_{\rm c}$, $L(T)/L_0$ broadly peaks around 0.3 K, and then
takes a distinct minimum at $T_{\rm min} \approx$ 0.095 K (upward arrow) below which $L(T)/L_0$ starts to rapidly rise upon cooling.
The maximum temperature roughly corresponds to a lower bound of $T$-linear $\rho$ observed at high temperature~\cite{custers300mK}.
In addition, below 0.3 K the magnetic susceptibility follows a Curie-Weiss law and the electronic specific heat deviates from the logarithmic behavior~\cite{custers,custers300mK}.
While the reason is unclear, the maximum is not clearly seen in the previous report with $B$ $\parallel$ $ab$~\cite{pfau}.
It is interesting to note that a similar upturn of $L(T)/L_0$ was observed in CeCoIn$_5$ with $q\parallel ab$
near the QCP, in which $L(T)/L_0$ rapidly increases to one, and thus the Wiedemann-Franz law was verified~\cite{paglione,seyfarth}.
The onset of the upturn was referred to $T_{\rm QP}$ as a characteristic energy scale below which well-defined quasiparticles are formed,
even though the standard FL behavior is absent; both $w(T)$ and $\rho(T)$ follow $T^{1.5}$ dependence below $T_{\rm QP}$~\cite{paglione}.
By analogy to CeCoIn$_5$, for YbRh$_2$Si$_2$ one can view the upturn of $L(T)/L_0$ below $T_{\rm min}$ as a signature of formation of 
well-defined quasiparticles, not due to the magnon contributions.

By increasing the field, $L(T)/L_0$ exhibits the following striking features:
1) $T_{\rm min}$ shifts to higher temperatures,
2) the width of the minimum becomes broader,
3) the slope of $L(T)/L_0$ below $T_{\rm min}$ gradually decreases, but
4) eventually $L(T)/L_0$ reaches one above 3 T within the measured temperature range.
This systematic increase of $L(T)/L_0$ below $T_{\rm min}$ towards unity by varying the fields further supports the notion that
the Wiedemann-Franz law is obeyed at the QCP.
The field variations of $T_{\rm min}$ together with $T_{\rm N}$ and $T_{\rm FL}$ derived from $\rho(T)$
are summarized in the inset of Fig.~\ref{fig.3}.
Note that the determined values of $T_{\rm N}$ and $T_{\rm FL}$ reproduce those of the previous report~\cite{gegenwart}, 
represented by the open symbols.
It is well demonstrated that 
at high fields $B \geq$ 1.5 T, $T_{\rm min}$ coincides with $T_{\rm FL}$.
However, as we approach the QCP,
$T_{\rm min}$ deviates from $T_{\rm FL}$ and remains finite, while $T_{\rm FL}$ vanishes, 
as similarly observed in CeCoIn$_5$~\cite{paglione},
suggesting that the quasiparticles remain intact at the QCP.

Finally, we turn to the field evolution of the Lorentz ratio $L(B)/L_0$.
Figure~\ref{fig.4} shows a $L(B)/L_0$ versus $B-B_{\rm c}$ plot derived from Fig.~\ref{fig.3} at finite temperatures
and from the extrapolated $T$ = 0 values in Fig.~\ref{fig.2}.
It is clearly seen that at high fields for $B \gg B_{\rm c}$,
$L(B)/L_0$ rapidly rises to one with lowering temperature, 
whereas on approaching the QCP it becomes significantly reduced even at 0.042 K due to the strong inelastic scattering from the critical fluctuations.
In the limit of $T$ = 0, however, $L(B)/L_0$ becomes one within the experimental error under the whole measured field for $B \geq B_{\rm c}$.
Again, this systematic evolution of $L(B)/L_0$ further supports the verification of the Wiedemann-Franz law at the QCP.

To date, experimental tests of the Wiedemann-Franz law at or near a QCP have been carried out for
CeCoIn$_5$~\cite{paglione,tanatar,seyfarth}, CeNi$_2$Ge$_2$~\cite{kambe}, and Sr$_3$Ru$_2$O$_7$~\cite{ronning}.
For the later two systems, the law was well verified within experimental accuracy.
In CeCoIn$_5$, however, a violation of the law was argued depending on the heat current directions
as a possible signature of anisotropic destruction of the Fermi surface~\cite{tanatar}.
Our results of the verification of the Wiedemann-Franz law at the field-tuned QCP of YbRh$_2$Si$_2$ rigorously points to no breakdown of the quasiparticle. 
In addition, there are no additional fermionic carriers of heat other than those which carry charge $e$.
This finding is obviously in contradiction to the Kondo breakdown scenario predicting the breakdown of the FL theory ~\cite{si,coleman} and
the upward violation of the Wiedemann-Franz law due to an existence of additional entropy carriers (spinons)~\cite{pepin}.
Instead, our results suggest that the quantum criticality of YbRh$_2$Si$_2$ can be described in the framework of a quasiparticle model,
such as quantum tricriticality scenario~\cite{imada} or a model of Zeeman-driven Lifshitz transition~\cite{vojta}.

In conclusion,
we have presented a set of thermal and charge conductivity data down to a very low temperature on the heavy-fermion compound YbRh$_2$Si$_2$. 
At finite temperature, we find the significant downward deviation from the Wiedemann-Franz law in the vicinity of the QCP, 
reflecting a dominant inelastic scattering contribution to the thermal conductivity.
In the limit $T\rightarrow$ 0, however, the Wiedemann-Franz law is found to be satisfied within experimental error at the QCP.
This finding explicitly points to no breakdown of the quasiparticle picture in YbRh$_2$Si$_2$, incompatible with the Kondo breakdown scenario.

This work is partially supported by Grants-in-Aids (Nos. 23340099, 23740263) 
for Scientific Research from the Japanese Society for the Promotion of Science (JSPS),
a Grant-in-Aid for Scientific Research on
Innovative Areas ``Heavy Electrons" (No. 20102006) of
The Ministry of Education, Culture, Sports, Science, and Technology, Japan (MEXT),
and a Grant-in-Aid for the Global COE Program from the MEXT through the Nanoscience and
Quantum Physics Project of the Tokyo Institute of Technology.

\bibliography{YRS}
\end{document}